\documentclass[9pt,conference]{IEEEtran}
\pdfoutput=1

\usepackage{cite}
\usepackage{amsmath,amssymb,amsfonts}
\usepackage{algorithmic}
\usepackage{graphicx}
\usepackage{textcomp}
\usepackage{xcolor}

\usepackage[caption=false,font=footnotesize]{subfig}

\usepackage{mleftright}

\usepackage{multirow}

\usepackage{threeparttable}

\usepackage{hyperref}

\renewcommand\hat{\widehat}

\def\N{\mathcal{N}}
\def\fN{\varphi}	
\def\FN{\phi}	

\newcommand{\Nms}[2][\mathbf{0}]{\N\mleft( #1,#2 \mathbf{I} \mright)}	
\def\NN{\Nms{}}										

\def\s{\sigma}
\def\ss{\s^{2}}

\def\eps{\epsilon}

\def\x{\mathbf{x}}
\def\y{\mathbf{y}}

\def\xh{\hat{\x}}
\def\xxt{\tilde{\x}}

\def\n{\mathbf{n}}

\def\ye{\y_e}
\def\ys{\y_s}

\def\U{U}
\def\Ui{\U^{-1}}

\def\ubk{u_{bk}}
\def\utbk{\ut_{bk}}

\def\u{\mathbf{u}}
\def\uh{\hat{\u}}

\def\ut{\tilde{u}}
\def\uut{\tilde{\u}}

\def\ub{\u_b}
\def\utb{\uut_b}

\def\ubm{\ub(m)}

\def\eb{e_b}

\def\ebm{\eb(m)}

\def\ie{y^e}
\def\is{y^s}
\def\iis{\mathbf{y}^s}

\def\ieb{\ie_b}
\def\iisb{\iis_b}
\def\isbk{\is_{bk}}

\def\m{\mathbf{m}}
\def\C{\mathbf{C}}

\newcommand{\pr}[1]{p\mleft( #1 \mright)}
\newcommand{\ppr}[1]{P\mleft( #1 \mright)}
\newcommand{\pc}[2]{\pr{#1 | #2}}
\newcommand{\ppc}[2]{\ppr{#1 | #2}}

\def\grad{\nabla}
\newcommand{\gx}[2]{\grad_{#1} \log \pr{ #1; #2}}
\newcommand{\gyx}[3]{\grad_{#2} \log \ppc{#1}{ #2; #3 }}

\newcommand{\St}[2]{S_{\theta}\mleft( #1 ; #2 \mright)}
\newcommand{\Sx}[2]{S_{#1}\mleft( #1; #2 \mright)}
\newcommand{\Sxy}[3]{S_{#1}\mleft( #1 | #2; #3 \mright)}
\newcommand{\Syx}[3]{S_{#2}\mleft( #1 | #2; #3 \mright)}

\begin{document}

\title{Audio Decoding by Inverse Problem Solving}

\author{
\IEEEauthorblockN{Pedro~J.~Villasana~T. \qquad Lars~Villemoes \qquad Janusz~Klejsa \qquad Per~Hedelin}
\IEEEauthorblockA{Dolby Sweden AB, Stockholm, Sweden}
}

\maketitle

\begin{abstract} 
We consider audio decoding as an inverse problem and solve it through diffusion posterior sampling. Explicit conditioning functions are developed for input signal measurements provided by an example of a transform domain perceptual audio codec. Viability is demonstrated by evaluating arbitrary pairings of a set of bitrates and task-agnostic prior models. For instance, we observe significant improvements on piano while maintaining speech performance when a speech model is replaced by a joint model trained on both speech and piano. With a more general music model, improved decoding compared to legacy methods is obtained for a broad range of content types and bitrates. The noisy mean model, underlying the proposed derivation of conditioning, enables a significant reduction of gradient evaluations for diffusion posterior sampling, compared to methods based on Tweedie's mean. Combining Tweedie's mean with our conditioning functions improves the objective performance. An audio demo is available at \href{https://dpscodec-demo.github.io/}{https://dpscodec-demo.github.io/}.
\end{abstract}

\begin{IEEEkeywords}
audio coding, diffusion posterior sampling, generative modeling, Langevin sampling
\end{IEEEkeywords}

\section{Introduction}\label{sec:intro}
Audio synthesis from features by finding consistency with measurements performed during encoding can be seen as inverse problem solving. Such a task has a long history in auditory research (e.g., in the context of synthesis by analysis \cite{slaney1995pattern}, texture synthesis \cite{mcdermott2009sound}, signal reconstruction based on perceptually motivated features \cite{villemoes2019learning}). Early approaches involved starting with Gaussian noise and performing iterative updates by filterbank processing or gradient descent to construct a signal that matches the encoder features. 

The significant advances in the field of neural audio coding rely on generative modeling to provide a signal prior \cite{Kleijn2018, Klejsa2019, davidson2023high}. In these works, supervised training of a generative model, conditioned on the signal representation provided by the encoder, needs to be performed for each target bitrate. In contrast to such an approach, in this paper, we investigate generative decoding of audio with an unsupervised model. We consider perceptual transform coding of audio and provide a proof of concept that facilitates signal reconstruction. 

Many general audio restoration problems can be formulated as inverse problems and then solved by diffusion posterior sampling. For example, source separation was considered in \cite{jayaram2021parallel, Villasana2023distribution}, bandwidth extension, declipping and inpainting in \cite{moliner2023solving, moliner2023blind}, and audio restoration in \cite{moliner2024diffusion, lemercier2024diffusion}. A promise of diffusion modeling is to realize approximations to Bayes rule for posterior sampling. This enables an unconditional prior model to be conditioned after its training with the aim of solving different inverse problems, simply described by their forward process. We consider the problem where audio encoding is the measurement and show that improved decoding can be achieved by task-agnostic diffusion prior models. Diffusion posterior sampling has so far been applied to image decoding encoded with a JPEG encoder \cite{kawar2022jpeg, song2023pseudoinverse}. 

A central challenge in the field is to design the conditioning score, see \cite{xu2024consistencymodeleffectiveposterior} and the references therein for the state-of-the-art on the topic. This requires a probabilistic model of the measurement of an unknown input signal given a noisy version of it. We propose a particularly straightforward approximation, the \emph{noisy mean model}. It allows us to compute explicit probabilities of the discrete measurements obtained by the considered audio encoder, namely signal membership in a bin of quantized values or envelopes. This method does not rely on the construction of a prior agnostic solution function to the inverse problem \cite{kawar2022jpeg, song2023pseudoinverse}. In contrast to the techniques \cite{moliner2023solving}, we have no tuning parameters in the conditioning and can omit an additional gradient propagation for each diffusion step. Still, we analyze the effect of applying approximations based on Tweedie's mean from \cite{song2023pseudoinverse} in lieu of the noisy mean model.

The paper is organized as follows. We describe the architecture of the perceptual audio codec used in this paper in Section \ref{sec:perceptual_audio_codec}. The inverse problem solution method is described in Section \ref{sec:audio_dec_inv}. Derivations for conditional scores are given in Section \ref{sec:cond_score}, and performance results are provided in Section \ref{sec:eval}.

\section{Perceptual Audio Codec}
\label{sec:perceptual_audio_codec}
\begin{figure}[h]
\centering
\includegraphics[scale=0.7]{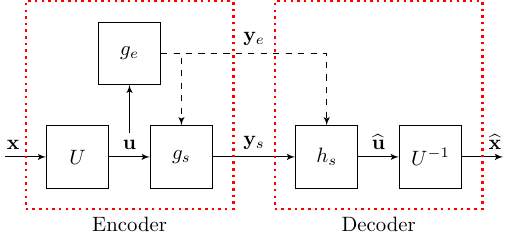}
\vspace*{-6pt}
\caption{A perceptual audio codec.} \label{fig:coder}
\vspace*{-12pt}
\end{figure}

We consider a transform coding architecture similar to \cite{fejgin2020sourcecoding}. The architecture, shown in Fig.~\ref{fig:coder}, employs a Modified Discrete Cosine Transform (MDCT), two-part description of signal (spectral envelope, MDCT lines), scalar quantization in a perceptual domain, and entropy coding. It facilitates operation with constrained bitrate.

\subsection{Encoder}
During encoding, a time-domain audio signal $\x$ is transformed to the MDCT domain using a stride of $L = 440$ samples, approximately corresponding to 20 ms at 22050~Hz sampling. Ignoring edge effects, this process is a unitary operator $\U$ with result $\u = \U(\x)$. For each frame $m$, the MDCT lines are banded into $B$ non-uniform, non-overlapping frequency bands. The envelope is then estimated as
\begin{align}\label{eq:env}
\ebm = \frac{1}{K_b} \| \ubm \|^2 
\end{align}
where $\ubm$ and $K_b$ are the coefficients and the number of MDCT lines, respectively, in the $b$-th band. The envelope values are then quantized with a $3$~dB stepsize. The resulting information indicates membership of envelope values $\ebm$ in a quantization interval $[e_L, e_H]$ for each band. We denote this joint information by $\ye = g_e(\u)$ in Fig.~\ref{fig:coder}.  

The MDCT lines $\ubm$ are then uniformly quantized with a stepsize approximately proportional to the fourth root of the midpoint of the envelope quantization interval, providing a perceptual shaping of quantization noise. The constant of proportionality in each frame represents an allocation level and is adjusted in 1.5~dB steps to meet the required bitrate. The resulting information indicates membership of each coordinate of $\ubm$ in a quantization bin of the form $[u_L, u_H]$. This joint information is denoted by $\ys = g_s(\u)$ in Fig.~\ref{fig:coder}.  

The indices of quantization bins and allocation levels are coded by standard lossless coding tools, namely differential coding and Huffman coding. The formulation of the inverse problem in Section~\ref{sec:audio_dec_inv} is based solely on the joint information in $\y = (\ye,\ys) = g(\x)$.

\subsection{Legacy Decoder}\label{sec:decoder}
For \emph{legacy decoding}, the first step is to reconstruct the MDCT lines $\uh$ from the information in $\ye$ and $\ys$ with \emph{deterministic reconstruction} points. This method is acceptable when the codec operates at high bitrate. However, when the bitrate is too low, the resulting decoded signal will have many artifacts such as spectral holes and loss of energy. To mitigate this, a perceptually tuned noise addition is performed for the three lowest allocation levels. This has the effect of filling spectral gaps with white noise shaped according to the envelope. The result in the transform domain is $\uh = h_s(\ye,\ys)$ in Fig.~\ref{fig:coder}, (omitting the allocation control).

Finally, the decoded MDCT lines are transformed back into the time domain by $\xh = \Ui(\uh)$. In the remainder of this paper, we refer to this decoding method as DEC.

\section{Audio Decoding as Inverse Problem Solving}\label{sec:audio_dec_inv}
In general, the optimal solution to fill-in the information lost during the encoding process should depend on the type of signals that is to be decoded. Here, we propose to reconstruct $\x$ by sampling from the conditional distribution $\pc{\x}{\y}$, given a signal prior model $\pr{\x}$. This results in a per-segment decoding method, as opposed to the per-frame operation of the legacy decoder.

Defining $\xxt = \x + \s\n$ with $\n \sim \NN$, we use Langevin sampling working with the log gradient of the posterior
\begin{align}\label{eq:scores}
\grad_{\xxt}\log \pc{\xxt}{\y}=\grad_{\xxt} \log \pr{\xxt} + \grad_{\xxt} \log \ppc{\y}{\xxt}.
\end{align}
As the encoded information is discrete, the last term of \eqref{eq:scores} is defined by a probability instead of a density. We indicate this by a capital letter $P$.

\subsection{Langevin Sampling}
The unconditional Langevin sampling algorithm introduced in \cite{song2019generative, song2020improved} generates a realization $\xxt$ iteratively as
\begin{align} \label{eq:lgv}
\xxt_{t_{i}} = \xxt_{t_{i-1}} + \alpha_{t_{i}} \gx{\xxt_{t_{i-1}}}{\s_{t_{i}}} + \beta_{t_{i}} \n_{t_{i}}
\end{align}
where $i = 1, \dotsc, I$ is the iteration number, and $t \in \mathcal{U}(0,1)$ describes the algorithmic time going from $t_{0} = 0$ to $t_{I} = 1$. Typically the noise steps $\s_{t_{0}} > \s_{t_{1}} > \dotsb > \s_{t_{I}}$ are chosen geometrically spaced, with $\s_{t_{0}} \gg \s_{x} \gg \s_{t_{I}}$ such that $\xxt_{t_{0}}$ is  indistinguishable from a realization sampled from $\Nms{\ss_{t_{0}}}$, and the amount of noise in $\xxt_{t_{I}}$ is perceptually negligible.

The selection of the hyperparameters $\alpha$ and $\beta$ in \eqref{eq:lgv} is a topic of research in itself and there are several choices available in the literature. In this paper we decided to use a mixture of the ``Annealed Langevin Sampling" method in \cite{song2019generative, song2020improved} and the ``Consistent Annealed Sampling" method in \cite{serra2021tuning} by choosing $\alpha_{t_{i}} = \eps \, \ss_{t_{i}}$ and $\beta_{t_{i}}^2 = 2 \alpha_{t_{i+1}}$. Therefore, the conditional Langevin sampling that generates $\xxt$ given $\y$ is implemented for $i = 1, \dotsc, I-1$ as
\begin{subequations} \label{eq:lgv_cond}
\begin{align}
\xxt_{t_{i}} = \xxt_{t_{i-1}} + \eps \, \ss_{t_{i}} \, \Sxy{\xxt_{t_{i-1}}}{\y}{\s_{t_{i}}} + \sqrt{2 \eps} \, \s_{t_{i+1}} \, \n_{t_{i}}
\end{align}
where the initial state $\xxt_{t_0}$ is a realization sampled from $\Nms{\ss_{t_0}}$, and for $i = I$
\begin{align} \label{eq:lgv_cond_I}
\xxt_{t_{I}} = \xxt_{t_{I-1}} + \eps \, \ss_{t_{I}} \, \Sxy{\xxt_{t_{I-1}}}{\y}{\s_{t_{I}}}.
\end{align}
\end{subequations}
Here
\begin{subequations}
\begin{align}
\Sxy{\xxt}{\y}{\s} = \Sx{\xxt}{\s} + \Syx{\y}{\xxt}{\s}
\end{align}
where
\begin{align}
\Sx{\xxt}{\s} &= \gx{\xxt}{\s}, \\
\Syx{\y}{\xxt}{\s} &= \gyx{\y}{\xxt}{\s}.
\end{align}
\end{subequations}

\subsection{Score Matching}
In practice, score matching models \cite{song2020score, pascual2023full, mariani2023multi} are used to estimate $\Sx{\xxt}{\s}$. These models typically follow the training procedure in \cite{song2019generative} that proposes score matching as a denoising algorithm that minimizes the loss function
\begin{align} \label{eq:loss}
\mathcal{L} = \mathbb{E}_{\x,\n,\s} \mleft[ \mleft\| \s \, \St{\x + \s \n}{\s} + \n \mright\|^2_2 \mright]
\end{align}
where $\St{\xxt}{\s}$ is the score model trying to match $\Sx{\xxt}{\s}$.

It is also possible to train conditional score models to match $\Sxy{\xxt}{\y}{\s}$. However, such a model becomes obsolete once the encoding $\y = g(\x)$ changes, e.g. when a particular element in Fig.~\ref{fig:coder} is replaced or when the coder runs at new bitrates. For this reason we propose to use a general unconditional score model $\St{\xxt}{\s}$ in conjunction with explicit conditioning functions $\Syx{\y}{\xxt}{\s}$.

\section{Conditional Scores}
\label{sec:cond_score}
To find the conditional score $\Sxy{\xxt}{\y}{\s}$ we need to find the distribution of $\y = g(\x)$ given $\xxt$. This problem is intractable in general, so we have to construct a model of $\x|\xxt$. Since $\xxt = \x+ \s \, \n$ we can write $\x = \xxt - \s\n$ and our aim is to find the distribution of $\y = g(\xxt - \s\n)$ given $\xxt$. Our heuristic approach is to treat $\xxt$ and $\n$ as independent. While this independence assumption is generally not true (except at the end of the Langevin sampling process), it has been used successfully in the past for tasks like source separation and inpainting resulting in working schemes \cite{jayaram2021parallel, Villasana2023distribution}. We name this approximation the \emph{noisy mean model}: 
\begin{align}\label{eq:tract}
\x|\xxt \sim \Nms[\xxt]{\ss}.
\end{align}
It facilitates the derivation of $\ppc{\y}{\xxt}$ for the encoding functions in Fig.~\ref{fig:coder}.

\subsection{Unitary Transforms}
When the transform $\U$ in Fig.~\ref{fig:coder} is unitary, the conditional scores can be computed in the transform domain and then transformed back into the time domain where the Langevin sampling operates,
\begin{align}\label{eq:score_trans}
\Syx{\y}{\xxt}{\s} = \mleft. \Ui \mleft( \Syx{\y}{\uut}{\s} \mright) \mright|_{\uut = \U(\xxt)}.
\end{align}

Furthermore, the \emph{noisy mean model} in the transform domain becomes
\begin{align} \label{eq:tract_trans}
\u | \uut \sim \Nms[\uut]{\ss}
\end{align}
where $\u = \U(\x)$ and $\uut = \U(\xxt)$. Therefore, one finds that the conditional scores in the transform domain can be computed in bands in a separable (block diagonal) fashion such that
\begin{align}
\Syx{\y}{\uut}{\s} = \mleft[ \Syx{\y_b}{\utb}{\s} \mright]_{1 \le b \le B}
\end{align}
where the indexing over the frame value $m$ has been dropped for clarity, as it becomes redundant under \eqref{eq:tract_trans}. We exploit this property to define conditional scores per band.

\begin{figure*}[t]
\vspace*{-12pt}
\centering
\subfloat{\includegraphics[width=0.25\textwidth]{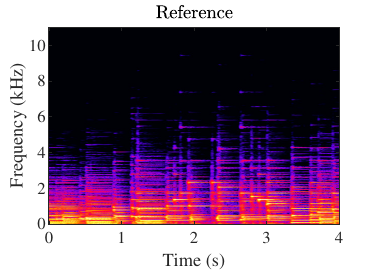}}
\subfloat{\includegraphics[width=0.25\textwidth]{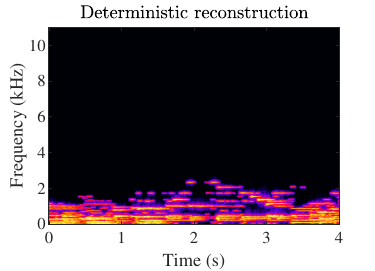}}
\subfloat{\includegraphics[width=0.25\textwidth]{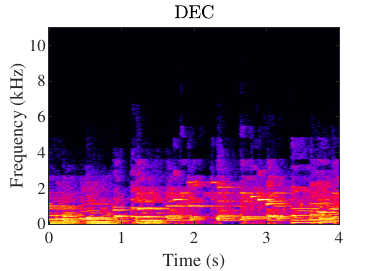}}
\subfloat{\includegraphics[width=0.25\textwidth]{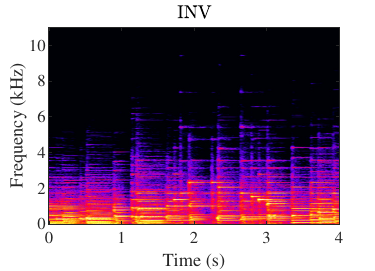}}
\vspace*{-6pt} \\
\caption{Spectrograms: (left) Input signal; (center-left) Deterministic reconstruction; (center-right) Legacy-decoded signal; (right) Diffusion-decoded signal.}
\label{fig:spectrograms}
\vspace*{-6pt}
\end{figure*}

\subsection{Band-wise Conditional Scores}
The conditioning information $\y_b$ for the $b$-th band indicates an envelope quantization interval $\ieb$ and sample quantization intervals $\iisb$. We introduce a second heuristic by treating these two types of conditions as independent, inspired by considering small values of $\s$ in \eqref{eq:tract_trans}. This leads to a simple superposition,
\begin{align} \label{eq:score_es}
\Syx{\y_b}{\utb}{\s} \approx \Syx{\ieb}{\utb}{\s} + \Syx{\iisb}{\utb}{\s}.
\end{align}
Both of these conditional scores can now be derived explicitly from the Gaussian assumption \eqref{eq:tract_trans}.

\subsubsection{Envelope Quantization Scores}
Assume that $\ieb$ describes the event $\eb(\ub) \in [ e_L(\ieb), e_H(\ieb)]$, where $\eb$ is defined as in \eqref{eq:env}. With $\xi_b(\ub) = K_b \, \eb(\ub) / \ss$, it follows from the \emph{noisy mean model} that
\begin{align*}
\xi_b(\ub) | \utb \sim {\chi'}^2_{K_b} \mleft( \xi_b(\utb) \mright)
\end{align*}
where ${\chi'}^2_{\nu}(\lambda)$ is the non-central Chi-squared distribution with $\nu$ degrees of freedom, and non-centrality parameter $\lambda$. With this convention, the conditional score for envelope quantization is found to be
\begin{align} \label{eq:score_e}
\Syx{\ieb}{\utb}{\s} = \frac{2}{\ss} \, q_{{\chi'}^2_{K_b}}( \xi_b( \utb ) ; \xi_{L}(\ieb) , \xi_{H}(\ieb) ) \, \utb 
\end{align}
where $\xi_{L}(\ieb) = K_b \, e_L(\ieb) /  \ss$, $\xi_{H}(\ieb) = K_b \, e_H(\ieb) / \ss$, and
\begin{align*}
q_{{\chi'}^2_{K}}( \lambda ; \xi_L, \xi_H) = \frac
{ f_{K+2}(\xi_L; \lambda) - f_{K+2}(\xi_H; \lambda) }
{  SF_{K}(\xi_L; \lambda) -  SF_{K}(\xi_H; \lambda) }.
\end{align*}
Here, $f_{\nu}(\xi; \lambda)$ and $SF_{\nu}(\xi; \lambda)$ are the PDF and survival function (complementary CDF) of $ {\chi'}^2_{\nu} \mleft( \lambda \mright)$, respectively.

\subsubsection{Sample Quantization Scores}
Assume that $\isbk$, the $k$-th element of $\iisb$, describes the event $\ubk \in [ u_L(\isbk) , u_H(\isbk) ]$, where $\ubk$ is the $k$-th sample of $\ub$. For the \emph{noisy mean model}, the conditional score for sample quantization is
\begin{align} \label{eq:score_sk}
\Syx{\isbk}{\utbk}{\s} = q_\s( \utbk ; u_L(\isbk) , u_H(\isbk) )
\end{align}
where $\utbk$ is the $k$-th sample of $\utb$, and
\begin{align*}
q_\s(n; n_L, n_H) = \frac{\fN_\s(n - n_L) - \fN_\s(n - n_H)}{\FN_\s(n - n_L) - \FN_\s(n - n_H)}.
\end{align*}
Here, $\fN_\s$ and $\FN_\s$ are the PDF and CDF of $\N(0,\ss)$, respectively.

The independence of sample events following from \eqref{eq:tract_trans} leads to a fully separable (diagonal) conditional score for the $b$-th band,
\begin{align} \label{eq:score_s}
\Syx{\iisb}{\utb}{\s} = \mleft[ \Syx{\isbk}{\utbk}{\s} \mright]_{1 \le k \le K_b}.
\end{align}

\subsection{Conditional Scores with Tweedie's Mean}
The \emph{noisy mean model} belongs to the family of Gaussian models $\x | \xxt \sim \N\mleft( \m(\xxt) , \C(\xxt) \mright)$. It is shown in \cite{boys2023tweedie} that the best approximation (in KL-divergence) of this type is achieved by using Tweedie's formula: $\m(\xxt) =\xxt + \ss \, \Sx{\xxt}{\s}$ and $\C(\xxt) = \ss \, \grad_{\xxt} \, \m(\xxt)$. By choosing $\m(\xxt)$ according to the \emph{noisy mean model}, the relationship between Tweedie's mean and covariance is preserved.

Tweedie's mean can be approximated as $\m(\xxt) \approx \xxt + \ss \, \St{\xxt}{\s}$, however Tweedie's moment involves computing the gradient of the score model. Furthermore, computing scores for $\y|\xxt$ involves computing gradients of both $\m(\xxt)$ and $\C(\xxt)$, and thus second derivatives of the score model. This was avoided in \cite{boys2023tweedie} by treating $\C(\xxt)$ as a constant.

Approximating Tweedie's moment as $\C(\xxt) \approx r^2 \mathbf{I}$, as in the $\Pi$GDM algorithm \cite{song2023pseudoinverse}, makes second derivatives through the score model unnecessary. The conditional scores for this approximation are then consistent with the formulas found in the current section as
\begin{align}
\Syx{\y}{\xxt}{\s} &= \mleft[ \grad_{\xxt} \, \m(\xxt) \mright] \, \mleft. \Ui \mleft( \Syx{\y}{\uut}{r(\s)} \mright) \mright|_{\uut = \U(\m(\xxt))}.
\end{align}

\begin{table}[b]
\vspace*{-12pt}
\centering
\caption{ViSQOL Scores for the Proposed Method (INV) for Different Combinations of Test Set, Model Prior, and Bitrate}
\label{table:experiments_with_models}
\begin{tabular}{|c|c|cccc|}
\hline
\textbf{Test}				& \textbf{Model}						& \multicolumn{4}{|c|}{\textbf{Bitrate [kb/s]}}					\\
\textbf{set}					& \textbf{prior}						& \textbf{8}		& \textbf{16}	& \textbf{24}	& \textbf{48}	\\
\hline
\multirow{4}{*}{Speech}		& Speech 							& \textbf{4.39}	& \textbf{4.53}	& \textbf{4.58}	& \textbf{4.62}	\\
							& Piano 								& 2.92			& 3.49			& 3.63			& 3.56			\\
							& Joint 								& \textbf{4.39}	& \textbf{4.53}	& 4.57			& \textbf{4.62}	\\
							& Music 								& 4.30			& 4.49			& 4.55			& 4.61			\\
\hline
\multirow{4}{*}{Piano}		& Speech							& 3.39			& 4.32			& 4.38			& 4.38			\\
							& Piano 								& \textbf{4.04}	& \textbf{4.48}	& \textbf{4.52}	& \textbf{4.52}	\\
							& Joint								& 3.96			& 4.44			& 4.48			& 4.48			\\
							& Music 								& 3.77			& 4.41			& 4.46			& 4.47			\\
\hline
\multirow{4}{*}{Critical}		& Speech							& 3.93			& 4.33			& 4.45			& 4.57			\\
							& Piano								& 2.57			& 3.00			& 3.31			& 3.65			\\
							& Joint								& 3.94			& 4.34			& 4.45			& 4.57			\\
							& Music								& \textbf{4.01}	& \textbf{4.35}	& \textbf{4.46}	& \textbf{4.58}	\\
\hline
\end{tabular}
\end{table}

\begin{table}[t]
\centering
\caption{ViSQOL Scores for DEC, INV, AAC, and Opus as Function of Bitrate for Different Test Sets}
\label{table:experiments_with_decoders}
\begin{tabular}{|c|c|cccc|}
\hline
\textbf{Test}				& \multirow{2}{*}{\textbf{Codec}}	& \multicolumn{4}{|c|}{\textbf{Bitrate [kb/s]}}					\\
\textbf{set}					&									& \textbf{8}		& \textbf{16}	& \textbf{24}	& \textbf{48}	\\
\hline
\multirow{4}{*}{Speech}		& DEC								& 4.21			& 4.42			& 4.49			& \textbf{4.64}	\\
							& INV$_\text{Speech}$				& \textbf{4.39}	& \textbf{4.53}	& \textbf{4.58}	& 4.62			\\
							& AAC								& 3.82			& 4.12			& 4.43			& 4.60			\\
							& Opus								& 2.96			& 3.80			& 4.00			& 4.05			\\
\hline
\multirow{4}{*}{Piano}		& DEC								& 3.72			& 4.47			& 4.50			& 4.51			\\
							& INV$_\text{Piano}$				& 4.04			& \textbf{4.48}	& 4.52			& 4.52			\\
							& AAC								& \textbf{4.11}	& 4.22			& \textbf{4.55}	& \textbf{4.72}	\\
							& Opus								& 3.17			& 3.93			& 4.10			& 4.25			\\
\hline
\multirow{4}{*}{Critical}		& DEC								& 3.93			& 4.27			& 4.38			& \textbf{4.58}	\\
							& INV$_\text{Music}$				& \textbf{4.01}	& \textbf{4.35}	& \textbf{4.46}	& \textbf{4.58}	\\
							& AAC								& 3.17			& 3.69			& 4.30			& 4.57			\\
							& Opus								& 2.35			& 3.86			& 4.12			& 4.29			\\
\hline
\end{tabular}
\end{table}

\begin{table}[t]
\centering
\begin{threeparttable}
\caption{ViSQOL Scores for INV with the Gaussian Model $\x|\xxt \sim \Nms[\m(\xxt)]{\ss}$}
\label{table:experiments_with_tweedie}
\begin{tabular}{|c|c|cccc|}
\hline
\textbf{Model}				& \textbf{Mean}\tnote{a}			& \multicolumn{4}{c|}{\textbf{Bitrate [kb/s]}}					\\
\textbf{prior}				& $\m(\xxt)$						& \textbf{8}		& \textbf{16}	& \textbf{24}	& \textbf{48}	\\
\hline
\multirow{2}{*}{Speech}		& Noisy								& 4.27			& 4.40			& 4.44			& 4.50			\\
							& Tweedie							& \textbf{4.31}	& \textbf{4.44}	& \textbf{4.49}	& \textbf{4.54}	\\
\hline
\multirow{2}{*}{Piano}		& Noisy								& 4.04			& \textbf{4.49}	& \textbf{4.52}	& \textbf{4.53}	\\
							& Tweedie							& \textbf{4.06}	& \textbf{4.49}	& \textbf{4.52}	& \textbf{4.53}	\\
\hline
\multirow{2}{*}{Music}		& Noisy								& 3.92			& 4.26			& 4.36			& 4.48			\\
							& Tweedie							& \textbf{3.97}	& \textbf{4.31}	& \textbf{4.42}	& \textbf{4.52}	\\
\hline
\end{tabular}
\begin{tablenotes}
\item[a] Noisy: $\m = \xxt$. Tweedie: $\m = \xxt + \ss \, \St{\xxt}{\s}$.
\end{tablenotes}
\end{threeparttable}
\vspace*{-10pt}
\end{table}

\section{Performance Analysis}\label{sec:eval}

\subsection{Experimental Setup}
We evaluate the proposed methodology on Speech (using the VCTK dataset \cite{yamagishi2019cstr}), Piano (using the Supra dataset \cite{shi2019supra}), and Music (internal) datasets. The datasets were resampled to 22050~Hz, each divided into 80\%-10\%-10\% subsets for training, validation, and testing. We also compiled a test set of critical signals derived from the ODAQ dataset \cite{torcoli2024odaq} (Critical). 

The unconditional models followed the DAG22 architecture in \cite{pascual2023full} for 22050~Hz audio signals. All models were trained to minimize the loss function \eqref{eq:loss} with $\s$ ranging from $\ss_{0}=0$~dB to $\ss_{1}=-90$~dB. The minimization was performed for 1M iterations on batches containing 16 audio signals of 8 seconds each, via gradient descent with the Adam optimizer, and a learning rate decreasing from $10^{-4}$ to $10^{-5}$ by means of cosine scheduling. For the Langevin sampling procedure in \eqref{eq:lgv_cond}, we set $\eps = 0.5$ and $I=1500$. An example reconstruction provided by the method is shown in Fig.~\ref{fig:spectrograms} for a piano signal.

\subsection{Objective Analysis}
We evaluated the robustness of the approach to model mismatch on the three datasets. Signals from the Speech, Piano, and Critical test sets were encoded at four bitrates. We performed decoding of these coded signals using the proposed method for diffusion posterior sampling (INV) with score models trained for different priors, including a joint Piano-Speech model by concatenating these datasets (Joint). Finally, an objective analysis of the reconstructed excerpts was performed using ViSQOL \cite{visqol}. The results of the evaluation are shown in Table~\ref{table:experiments_with_models}. For decoders with prior models matching the test signals, the scores clearly indicate quality reconstruction increasing with the bitrate. This observation extends to the scheme with the Joint model as a prior. Mismatched models, as expected, achieve worse performance. The Piano model was particularly sensitive to test mismatch. It produced artifacts for some of the test items, thus resulting in lower average scores.

In a second experiment, we evaluated the performance of the proposed decoding method as a function of bitrate, comparing it to the legacy decoding method (DEC) for our encoder, as well as to AAC (32~kHz tuning) \cite{Bosi1997}, and Opus (operating with a constant bitrate constraint) \cite{Valin2012}. The scores, collected in Table~\ref{table:experiments_with_decoders}, show a robust performance improvement of INV over DEC. They also indicate competitive performance with respect to the anchor codecs.

Finally, in a third experiment we compared objective scores when using the \emph{noisy mean model} vs. using Tweedie's mean. Reconstruction of some excerpts turned out numerically unstable when using Tweedie's mean, especially as $\s$ became very small. Therefore, Langevin sampling  for this experiment was performed with $I = 1250$ and $\ss_{t_{I}} = -75$~dB (for both the \emph{noisy mean model} and Tweedie's mean). The scores can be seen in Table~\ref{table:experiments_with_tweedie}, showing a modest improvement when Tweedie's mean was used. However, the use of Tweedie's mean requires computing gradients of the score model at every iteration, which increased the sampling time by 1.5x and the memory usuage by 2x. We also tested changing the covariance to that of the $\Pi$GDM approach \cite{song2023pseudoinverse} $\C = r^2 \mathbf{I}$ with $r^2 = \ss / (1 + \ss)$ but this provided no significant improvement.

\begin{figure}[t]
\centering
\includegraphics[width=0.9\linewidth]{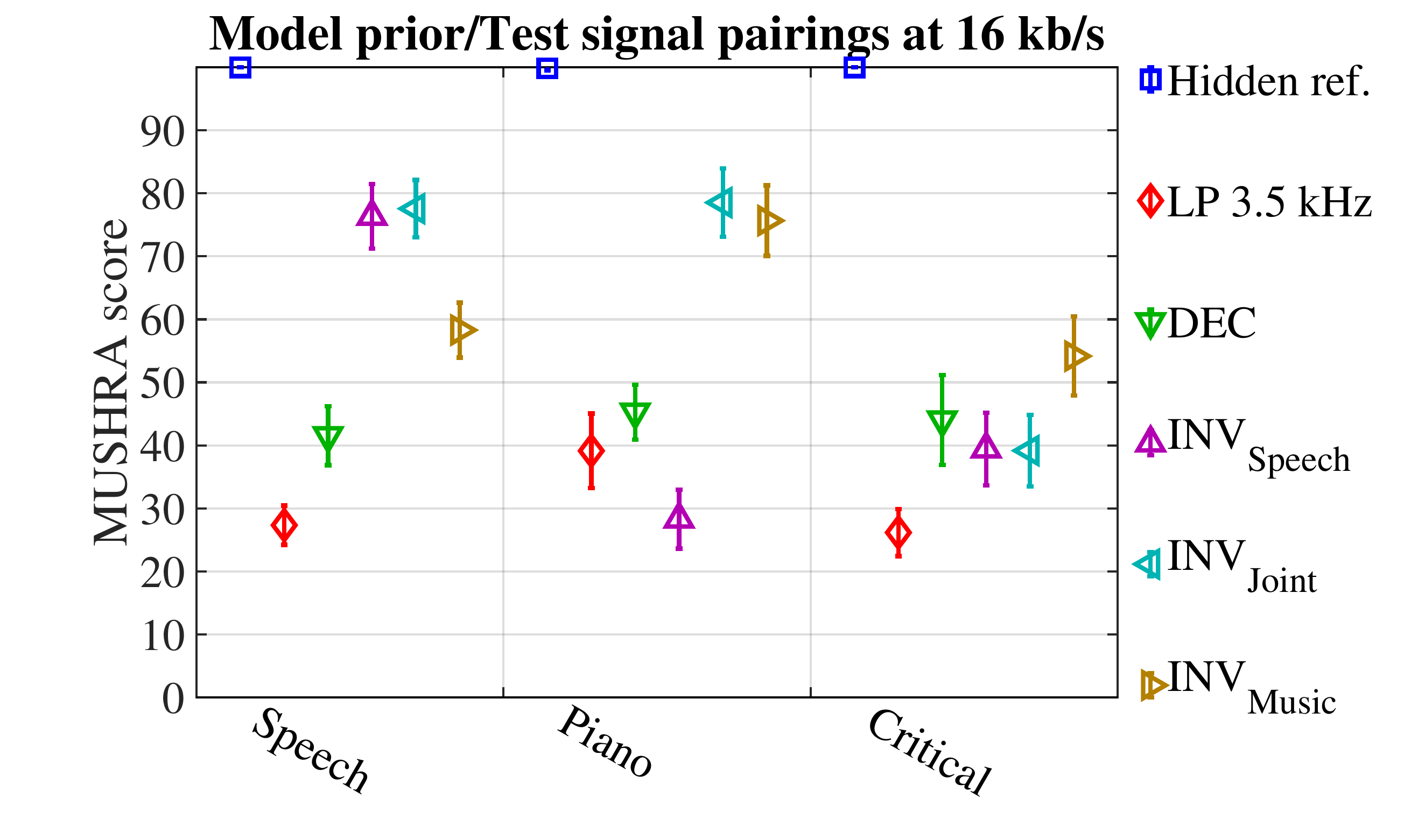}
\vspace*{-16pt}
\caption{Results of a MUSHRA-like listening test at 16~kb/s (3 items per category, 10 listeners, Student-t 95\% confidence intervals).}
\label{fig:mushra_test_model}
\vspace*{-6pt}
\end{figure}

\begin{figure}[t]
\centering
\includegraphics[width=0.9\linewidth]{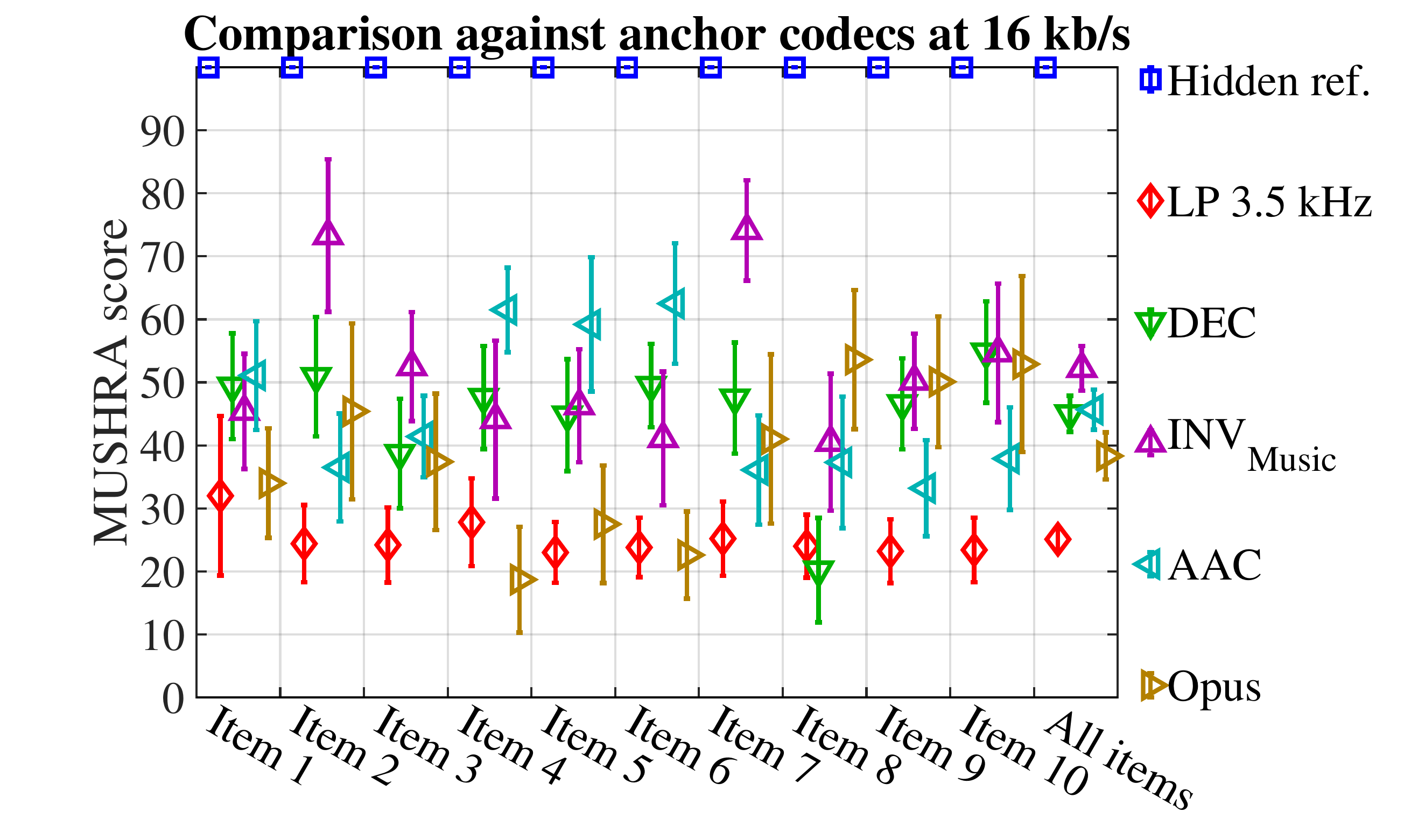}
\vspace*{-12pt}
\caption{Results of a MUSHRA-like listening test at 16~kb/s on the Critical test set (10 listeners, Student-t 95\% confidence intervals).}
\label{fig:mushra_test_codec}
\vspace*{-12pt}
\end{figure}

\subsection{Listening Tests}
Two MUSHRA-like listening tests \cite{bs.1534-3} were conducted to validate the provided ViSQOL scores and to compare the legacy decoding method (DEC) against the proposed diffusion-based decoding algorithm (INV). The audio signals of both tests were encoded at 16~kb/s. Also included in the tests were a hidden reference and a 3.5~kHz low-pass anchor.

The first listening test evaluated the performance of INV for arbitrary pairings of prior models to signal categories. The test contained 3 randomly selected items from each of the Speech, Piano and Critical test sets, for a total of 9 items. The conditions in the test included three instances of INV employing unconditional models trained on the Speech, Joint, and Music training sets. The per-category results of this listening test are shown in Fig.~\ref{fig:mushra_test_model}. There were 10 expert listeners participating in this test. It can be seen that the results corroborate the objective analysis shown in Table~\ref{table:experiments_with_models}.

For the second listening test, 10 critical signals were selected from the Critical test set. The INV$_\text{Music}$ model was chosen for this test. Also in the test were included items coded at 16~kb/s with AAC, and Opus. There were 10 expert listeners participating in the listening test. The results of the test are presented in Fig.~\ref{fig:mushra_test_codec}. It can be seen that INV$_\text{Music}$ provided a significant improvement over DEC and anchor codecs (as indicated by the ViSQOL scores shown in Table~\ref{table:experiments_with_decoders}).

\section{Conclusion}
By constructing a first proof of concept, we provide evidence that audio decoding can be treated as an inverse encoding problem and solved by posterior sampling. Further advances are to be expected on both generative models and conditioning algorithms. The approach enables decoders to draw on a wide selection of prior models for quality improvements at low bit rates. At the same time, control of the content intention can be maintained through perceptual encoder design.

\bibliographystyle{IEEEtran}
\bibliography{IEEEabrv,main}

\begin{thebibliography}{10}
\providecommand{\url}[1]{#1}
\csname url@samestyle\endcsname
\providecommand{\newblock}{\relax}
\providecommand{\bibinfo}[2]{#2}
\providecommand{\BIBentrySTDinterwordspacing}{\spaceskip=0pt\relax}
\providecommand{\BIBentryALTinterwordstretchfactor}{4}
\providecommand{\BIBentryALTinterwordspacing}{\spaceskip=\fontdimen2\font plus
\BIBentryALTinterwordstretchfactor\fontdimen3\font minus
  \fontdimen4\font\relax}
\providecommand{\BIBforeignlanguage}[2]{{%
\expandafter\ifx\csname l@#1\endcsname\relax
\typeout{** WARNING: IEEEtran.bst: No hyphenation pattern has been}%
\typeout{** loaded for the language `#1'. Using the pattern for}%
\typeout{** the default language instead.}%
\else
\language=\csname l@#1\endcsname
\fi
#2}}
\providecommand{\BIBdecl}{\relax}
\BIBdecl

\bibitem{slaney1995pattern}
M.~Slaney, ``Pattern playback from 1950 to 1995,'' in \emph{1995 IEEE
  International Conference on Systems, Man and Cybernetics. Intelligent Systems
  for the 21st Century}, vol.~4.\hskip 1em plus 0.5em minus 0.4em\relax IEEE,
  1995, pp. 3519--3524.

\bibitem{mcdermott2009sound}
J.~H. McDermott, A.~J. Oxenham, and E.~P. Simoncelli, ``Sound texture synthesis
  via filter statistics,'' in \emph{2009 IEEE Workshop on Applications of
  Signal Processing to Audio and Acoustics}.\hskip 1em plus 0.5em minus
  0.4em\relax IEEE, 2009, pp. 297--300.

\bibitem{villemoes2019learning}
L.~Villemoes, A.~Biswas, H.~Purnhagen, and H.-M. Lehtonen, ``Learning about
  perception of temporal fine structure by building audio codecs,'' in
  \emph{Proceedings of the International Symposium on Auditory and Audiological
  Research}, vol.~7, 2019, pp. 141--148.

\bibitem{Kleijn2018}
W.~B. Kleijn, F.~S.~C. Lim, A.~Luebs, J.~Skoglund, F.~Stimberg, Q.~Wang, and
  T.~C. Walters, ``{WaveNet} based low rate speech coding,'' in \emph{2018 IEEE
  International Conference on Acoustics, Speech and Signal Processing
  (ICASSP)}, April 2018, pp. 676--680.

\bibitem{Klejsa2019}
J.~Klejsa, P.~Hedelin, C.~Zhou, R.~Fejgin, and L.~Villemoes, ``High-quality
  speech coding with {sampleRNN},'' in \emph{ICASSP 2019-2019 IEEE
  International Conference on Acoustics, Speech and Signal Processing
  (ICASSP)}.\hskip 1em plus 0.5em minus 0.4em\relax IEEE, 2019, pp. 7155--7159.

\bibitem{davidson2023high}
G.~Davidson, M.~Vinton, P.~Ekstrand, C.~Zhou, L.~Villemoes, and L.~Lu, ``High
  quality audio coding with {MDCTNet},'' in \emph{ICASSP 2023-2023 IEEE
  International Conference on Acoustics, Speech and Signal Processing
  (ICASSP)}.\hskip 1em plus 0.5em minus 0.4em\relax IEEE, 2023, pp. 1--5.

\bibitem{jayaram2021parallel}
V.~Jayaram and J.~Thickstun, ``Parallel and flexible sampling from
  autoregressive models via langevin dynamics,'' in \emph{International
  Conference on Machine Learning}.\hskip 1em plus 0.5em minus 0.4em\relax PMLR,
  2021, pp. 4807--4818.

\bibitem{Villasana2023distribution}
P.~J. Villasana~T, J.~Klejsa, L.~Villemoes, P.~Hedelin \emph{et~al.},
  ``Distribution preserving source separation with time frequency predictive
  models,'' in \emph{2023 31st European Signal Processing Conference
  (EUSIPCO)}.\hskip 1em plus 0.5em minus 0.4em\relax IEEE, 2023, pp. 46--50.

\bibitem{moliner2023solving}
E.~Moliner, J.~Lehtinen, and V.~V{\"a}lim{\"a}ki, ``Solving audio inverse
  problems with a diffusion model,'' in \emph{ICASSP 2023-2023 IEEE
  International Conference on Acoustics, Speech and Signal Processing
  (ICASSP)}.\hskip 1em plus 0.5em minus 0.4em\relax IEEE, 2023, pp. 1--5.

\bibitem{moliner2023blind}
E.~Moliner, F.~Elvander, and V.~V{\"a}lim{\"a}ki, ``Blind audio bandwidth
  extension: A diffusion-based zero-shot approach,'' \emph{arXiv preprint
  arXiv:2306.01433}, 2023.

\bibitem{moliner2024diffusion}
E.~Moliner, M.~Turunen, F.~Elvander, and V.~V{\"a}lim{\"a}ki, ``A
  diffusion-based generative equalizer for music restoration,'' \emph{arXiv
  preprint arXiv:2403.18636}, 2024.

\bibitem{lemercier2024diffusion}
J.-M. Lemercier, J.~Richter, S.~Welker, E.~Moliner, V.~V{\"a}lim{\"a}ki, and
  T.~Gerkmann, ``Diffusion models for audio restoration,'' \emph{arXiv preprint
  arXiv:2402.09821}, 2024.

\bibitem{kawar2022jpeg}
B.~Kawar, J.~Song, S.~Ermon, and M.~Elad, ``{JPEG} artifact correction using
  denoising diffusion restoration models,'' \emph{arXiv preprint
  arXiv:2209.11888}, 2022.

\bibitem{song2023pseudoinverse}
J.~Song, A.~Vahdat, M.~Mardani, and J.~Kautz, ``Pseudoinverse-guided diffusion
  models for inverse problems,'' in \emph{International Conference on Learning
  Representations}, 2023.

\bibitem{xu2024consistencymodeleffectiveposterior}
\BIBentryALTinterwordspacing
T.~Xu, Z.~Zhu, J.~Li, D.~He, Y.~Wang, M.~Sun, L.~Li, H.~Qin, Y.~Wang, J.~Liu,
  and Y.-Q. Zhang, ``Consistency model is an effective posterior sample
  approximation for diffusion inverse solvers,'' \emph{arXiv preprint
  arXiv:2403.12063}, 2024. [Online]. Available:
  \url{https://arxiv.org/abs/2403.12063}
\BIBentrySTDinterwordspacing

\bibitem{fejgin2020sourcecoding}
R.~Fejgin, J.~Klejsa, L.~Villemoes, and C.~Zhou, ``Source coding of audio
  signals with a generative model,'' in \emph{ICASSP 2020 - 2020 IEEE
  International Conference on Acoustics, Speech and Signal Processing
  (ICASSP)}.\hskip 1em plus 0.5em minus 0.4em\relax IEEE, 2020, pp. 341--345.

\bibitem{song2019generative}
Y.~Song and S.~Ermon, ``Generative modeling by estimating gradients of the data
  distribution,'' \emph{Advances in neural information processing systems},
  vol.~32, 2019.

\bibitem{song2020improved}
------, ``Improved techniques for training score-based generative models,'' in
  \emph{Proceedings of the 34th International Conference on Neural Information
  Processing Systems}, ser. NIPS '20.\hskip 1em plus 0.5em minus 0.4em\relax
  Red Hook, NY, USA: Curran Associates Inc., 2020.

\bibitem{serra2021tuning}
J.~Serr{\`a}, S.~Pascual, and J.~Pons, ``On tuning consistent annealed sampling
  for denoising score matching,'' \emph{arXiv preprint arXiv:2104.03725}, 2021.

\bibitem{song2020score}
Y.~Song, J.~Sohl-Dickstein, D.~P. Kingma, A.~Kumar, S.~Ermon, and B.~Poole,
  ``Score-based generative modeling through stochastic differential
  equations,'' \emph{arXiv preprint arXiv:2011.13456}, 2020.

\bibitem{pascual2023full}
S.~Pascual, G.~Bhattacharya, C.~Yeh, J.~Pons, and J.~Serrà, ``Full-band
  general audio synthesis with score-based diffusion,'' in \emph{ICASSP 2023 -
  2023 IEEE International Conference on Acoustics, Speech and Signal Processing
  (ICASSP)}, 2023, pp. 1--5.

\bibitem{mariani2023multi}
G.~Mariani, I.~Tallini, E.~Postolache, M.~Mancusi, L.~Cosmo, and E.~Rodol{\`a},
  ``Multi-source diffusion models for simultaneous music generation and
  separation,'' \emph{arXiv preprint arXiv:2302.02257}, 2023.

\bibitem{boys2023tweedie}
\BIBentryALTinterwordspacing
B.~Boys, M.~Girolami, J.~Pidstrigach, S.~Reich, A.~Mosca, and O.~D. Akyildiz,
  ``Tweedie moment projected diffusions for inverse problems,'' 2023. [Online].
  Available: \url{https://arxiv.org/abs/2310.06721}
\BIBentrySTDinterwordspacing

\bibitem{yamagishi2019cstr}
J.~Yamagishi, C.~Veaux, K.~MacDonald \emph{et~al.}, ``{CSTR VCTK Corpus}:
  English multi-speaker corpus for {CSTR} voice cloning toolkit (version
  0.92),'' \emph{University of Edinburgh. The Centre for Speech Technology
  Research (CSTR)}, 2019.

\bibitem{shi2019supra}
Z.~Shi, C.~Sapp, K.~Arul, J.~McBride, and J.~O. Smith~III, ``{SUPRA}:
  Digitizing the {Stanford University Piano Roll Archive},'' in \emph{ISMIR},
  2019, pp. 517--523.

\bibitem{torcoli2024odaq}
M.~Torcoli, C.-W. Wu, S.~Dick, P.~A. Williams, M.~M. Halimeh, W.~Wolcott, and
  E.~A. Habets, ``{ODAQ}: Open dataset of audio quality,'' in \emph{ICASSP
  2024-2024 IEEE International Conference on Acoustics, Speech and Signal
  Processing (ICASSP)}.\hskip 1em plus 0.5em minus 0.4em\relax IEEE, 2024, pp.
  836--840.

\bibitem{visqol}
Google, ``{ViSQOL},'' https://github.com/google/visqol, 2022.

\bibitem{Bosi1997}
M.~Bosi, K.~Brandenburg, S.~Quackenbush, L.~Fielder, K.~Akagiri, H.~Fuchs, and
  M.~Dietz, ``{ISO/IEC MPEG-2 Advanced Audio Coding},'' \emph{Journal of the
  Audio Engineering Society}, vol.~45, no.~10, pp. 789--814, 1997.

\bibitem{Valin2012}
\BIBentryALTinterwordspacing
J.-M. Valin, K.~Vos, and T.~Terriberry, ``Definition of the {Opus} audio
  codec,'' \emph{{RFC 6716, IETF}}, 2012. [Online]. Available:
  \url{https://tools.ietf.org/html/rfc6716}
\BIBentrySTDinterwordspacing

\bibitem{bs.1534-3}
``Method for the subjective assessment of intermediate quality levels of coding
  systems,'' Recommendation ITU-R BS.1534-3, Oct. 2015.

\end{thebibliography}

\end{document}